\documentclass[12pt]{iopart}

\input{epsf}

\begin{document}

\title[Kinetic glass transition]{Kinetic glass transition
}

\author{        Mauro Sellitto}

\address{       Laboratoire de Physique 
                de l'\'Ecole Normale Sup\'erieure de Lyon \\
                46 All\'ee d'Italie, 69007 Lyon, France \\
} 
\begin{abstract}
        Kinetic lattice-gas models display fragile glass behavior, 
        in spite of their trivial Gibbs-Boltzmann measure. 
        This suggests that the nature of glass transition might be, at least 
        in some cases, understood in purely kinetic or dynamical terms.
\end{abstract}


\maketitle


\section{Introduction}

A long debated problem in glass physics concerns the nature of the dynamical 
ergodicity breaking and its relation with the
existence of an underlying equilibrium phase transition~\cite{Kauz,GiMa}.
In mode-coupling theory the glass transition appears as a purely 
dynamic effect due to an instability of the equation governing the 
time correlation of density fluctuations~\cite{Leu,BeGoSj}. 
In particular, mean-field disordered models of structural glasses 
show that glassy features are associated to a rugged free energy landscape
and that the origin of the dynamical transition is the existence 
of a large number of metastable states which trap the system for an
infinite time~\cite{KW,KT}.
On the other hand, the lifetime of metastable states in finite dimensional
short range models is finite, since it is always possible to nucleate, by a
thermally activated process, a droplet of the stable phase.
Therefore the dynamical transition appear as an artifact
of the mean-field approximation, and in real glasses this transition
would be just a {\it finite-time\/} kinetic effect, at least on time 
scales much smaller than the lifetime of metastable states.
Recently, the close connection between the non-trivial structure of Gibbs 
equilibrium states and the appearance of a persistent glassy dynamics has 
been established for a certain class of systems~\cite{FrMePaPe}. 
However, since the dynamical universality classes are smaller than the 
static ones~\cite{Ka,HoHa}, and since salient features of glasses are 
essentially of dynamical nature~\cite{Fred}, it is important to understand 
to what extent glassy behaviour depends on  the details of microscopic 
kinetics.
A generic microscopic mechanism leading to slow relaxation phenomena
was suggested some time ago by Fredrickson and Andersen~\cite{FrAn}.
It is based on kinetic rules involving only a selection of the
possible configuration changes compatibly with the detailed balance
and the Boltzmann distribution.
A kinetic rule can be so effective that there is no need to
introduce an energetic interaction between the particles.
Although they are physically motivated (e.g. by the cage effect mechanism)
these kinetic models are not intended to describe the realistic dynamics 
of glasses, but rather to show that the glass transition could be, at least in 
principle, a purely kinetic or dynamical phenomenon.
Taking advantage of this idea we have explored the limit case of
a three dimensional lattice gas model defined only by short-range kinetic
constraints and by a trivial equilibrium measure~\cite{KoAn}.
Remarkably, this finite dimensional model exhibits a {\it fragile} glass 
behavior unrelated to the existence of a thermodynamic phase 
transition~\cite{KoAn,KuPeSe,PeSe}
(for another case and the related experimental situation, 
see~\cite{ReLoBa,ObKoPeRu} and~\cite{Ji}).
It provides a simple example of how the distinction between the {\it ideal}
(static or dynamic) and  the {\it laboratory} (i.e. kinetic) 
glass transition can be very subtle and elusive.
In the following we present some numerical results showing that this 
model reproduces qualitatively some aspects of the glassy phenomenology, 
such as history dependence, 
irreversibility effects, power-law approach to the asymptotic state, 
and simple aging behavior. Some related works on constrained 
lattice-gas models are~\cite{EiJa,FoRi,GrPiGr,ScTr}.

\section{The model}

Our starting point is a kinetic lattice-gas model introduced 
by Kob and Andersen~\cite{KoAn}.
The system consists of $N$ particles in a cubic lattice of size $L^{3}$, 
with periodic boundary conditions.
There can be at most one particle per site. 
Apart from this hard-core constraint there are no other static 
interactions among the particles.
At each time step a particle and one of its neighbouring sites are 
chosen at random. 
The particle moves if the three following conditions are all met:
\begin{enumerate}
\item the neighbouring site is empty;
\item the particle has less than $m$ nearest neighbours;
\item the particle will have less than $m$ nearest neighbours after 
it has moved.
\end{enumerate}
The rule is symmetric in time, detailed balance is satisfied and the 
allowed configurations have the same statistical weight in equilibrium.
Significant results are obtained when the value of $m$ is set to $4$.
With this simple definition one can proceed to study the dynamical 
behavior of the model at equilibrium. 
One observes that the dynamics becomes slower and slower as the 
particle density $\rho$ increases; in particular, the diffusion 
coefficient of the particles, $D$, vanishes as
the density $\rho$ approaches the
critical value $\rho_{\rm c}\simeq 0.88$, with a power law 
\begin{eqnarray}
D(\rho) & \sim & (\rho_{\rm c}-\rho)^{\phi}, 
\end{eqnarray}
with an exponent $\phi \simeq 3.1$~\cite{KoAn}.
Since we are interested in the dynamical approach to the putative 
equilibrium state we allow the system to exchange particles with a
reservoir characterized by a chemical potential $\mu$.
Therefore, we alternate the ordinary diffusion sweeps with
sweeps of creation/destruction of particles on a single layer 
with the following Monte Carlo rule: we randomly choose a site on 
the layer; if it is empty, we add a new particle; otherwise we 
remove the old particle with probability $\mbox{e}^{-\beta \mu}$.
The number of particles is no longer fixed and the external control 
parameter is $1/\mu$, which plays the role of the temperature.
The equilibrium equation of state $\rho=\rho_{\rm eq}(\mu)$ is then trivially
calculated. There is therefore a critical value $\mu_{\rm c}$ of $\mu$
defined by $\rho_{\rm eq}(\mu_{\rm c})=\rho_{\rm c}$ 
corresponding to the ideal glass transition of the model.
In this way we can prepare the system in a non equilibrium state
by a process analogous to a quench,
which is represented by a jump in  $1/\mu$ from above to below
$1/\mu_{\rm c}$. Or, we can let $1/\mu$ decrease or increase smoothly
like in cooling or heating experiments.
The situation becomes analogous to the {\it canonic\/} case in 
which one controls the temperature, and the energy endeavors to reach  
its equilibrium value.

\section{Thermodynamics}

Before to study the non-equilibrium regime let us
consider the static properties.
The point is relevant for the question of whether the possible ideal glass
transition is purely dynamical or is a consequence of an equilibrium 
transition of some sort.
The Hamiltonian of the model is
\begin{eqnarray}
{\cal H} &=& - \mu \, \sum_{i=1}^N n_i  \,\,,
\end{eqnarray}
where $n_i=0,1$ are occupation site variables and $\mu$ is the chemical
potential. The corresponding partition function, for a system of volume
$V=L^3$,  
\begin{eqnarray}
Z &=&  \left( 1+{\mbox{e}}^{\beta \mu} \right)^V \,\,,
\label{ZZ}
\end{eqnarray}
would describe correctly the thermodynamics of the system provided
that the measure of configurations made inaccessible by the kinetic 
constraints vanishes in the thermodynamic limit.             
It is possible to convince oneself that the kinetic rules, which
satisfy detailed balance, allow an initially empty lattice to be 
progressively filled in, leaving only O($1/L$) empty sites per unit volume.
Indeed, it is always possible to 
find a path connecting almost any two allowed configurations, 
if necessary by letting the particles escape one by one by the way they 
got in.
Therefore the Markov process generated by the dynamical evolution rule is 
irreducible on the full manifold of particles configurations and the
static properties of the model are described by the~(\ref{ZZ}). 
In particular the state equation and the entropy
are respectively given by: 
\begin{eqnarray}
 \rho &=& 1/(1+{\rm e}^{-\beta \mu}) \,\,,  \\
S  &=& - \rho \log \rho -(1-\rho) \log (1-\rho) \,\,.
\end{eqnarray}
Since the static properties of the system are regular
as a function of the density or the chemical potential, 
the possible ideal glass transition should appear as a 
purely dynamical effect.                 
The critical value of $\mu$ and  $S$ corresponding to the threshold
density $\rho_{\rm c}$, can be estimated from the previous equations and 
they are given by
\begin{eqnarray}
\mu_{\rm c} \simeq  2.0 \,\,,  \qquad S_{\rm c} \simeq 0.36 \,\,.
\end{eqnarray}

\section{History dependence}
 
\begin{figure}
\begin{center}
\leavevmode
\epsfysize=230pt{\epsffile{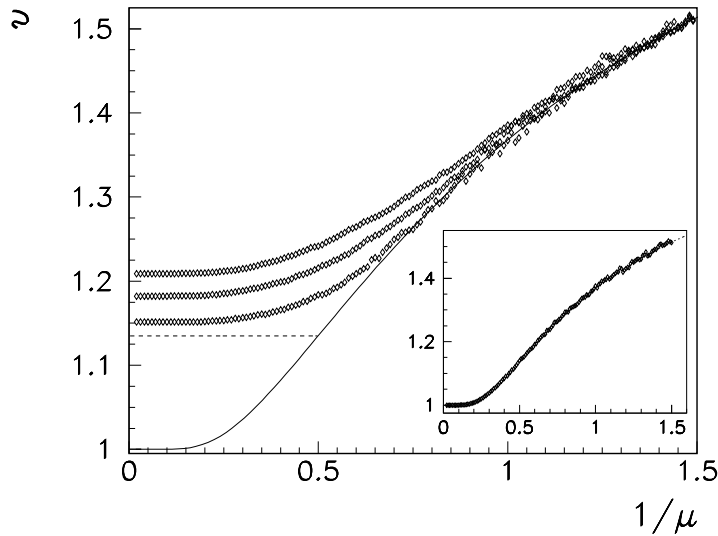}}
\end{center}
\protect\caption{ Compression experiment. 
The smooth curve is the equilibrium state equation $v=v(1/\mu)$; 
while the dashed line is the critical value
of specific volume  $v_{\rm c}\simeq 1.13$.
The compression rates are 
(from top to bottom) $r= 1/3 \cdot 10^{-4}, 10^{-5},10^{-6} $ in unit 
of [$\mu \cdot {\rm MC \, sweep}$]$^{-1}$. 
Inset:
the same experiment in a system without kinetic constraints;
here the rate is $r=10^{-4}$.
\protect\label{annealing} }
\end{figure}

A first insight into the nature of the relaxational processes
can be gained by studying the behavior of one-time observables 
(energy, specific volume etc.) in a slow annealing procedure. 
We consider a compression experiment in which the inverse
chemical  potential of the reservoir, $1/\mu$, is slowly decreased at fixed 
rate from a value corresponding to a low density equilibrium 
configuration up to zero.                                       
The simulation results presented in the following refer to a system 
of size $20^3$.
In Figure~\ref{annealing} the numerical results of the specific volume 
$v=1/\rho$ vs. $1/\mu$ are compared, for several annealing rates, $r$, with 
the equilibrium state equation of the system (the smooth curve).        
In close resemblance with the behavior of real glasses
these curves exhibits the characteristic annealing dependence 
of one-time observables:
after a certain value of the inverse chemical potential, $1/\mu_{\rm g}(r)$, 
is reached the dynamics become so sluggish that the system is no longer 
able to follow the annealing procedure;
the faster the compression, the sooner the system falls off
equilibrium.
The limit value of $v$ reaches a plateau that depends on 
the compression rate and never seems to cross the critical value 
$v_{\rm c}=1/\rho_{\rm c}$ (the horizontal dashed line). 
In the inset of fig.~\ref{annealing} we also show the same plot 
for a compression experiment, but this time by removing the dynamical 
constraints.
We see that the ordinary lattice gas has no problem in equilibrating 
at each value of chemical potential $\mu$; 
therefore our ``experimental'' setup (the way in which the particles
reservoir and its connection with the system is realized) provide a 
suitable representation of the equilibrium properties of the model.
\begin{figure}
\begin{center}
\leavevmode
\epsfysize=230pt{\epsffile{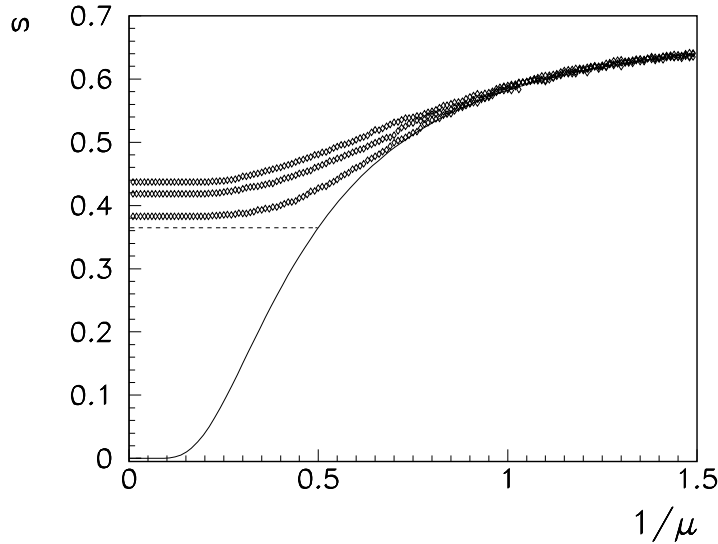}}
\caption{Calorimetric entropy as obtained by integration 
of compression experiment data.
The smooth curve is the equilibrium entropy $S=S(1/\mu)$ while 
the dashed line is the critical value of entropy 
$S_{\rm c} \simeq 0.365 $.
The compression rates are (from top to bottom) $r=1/3 \cdot 10^{-4}, 
10^{-5} ,10^{-6}$.  }
\label{entropy}
\end{center}
\end{figure}
\subsubsection*{Kauzmann's paradox.} 
\begin{figure}
\begin{center}
\leavevmode
\epsfysize=230pt{\epsffile{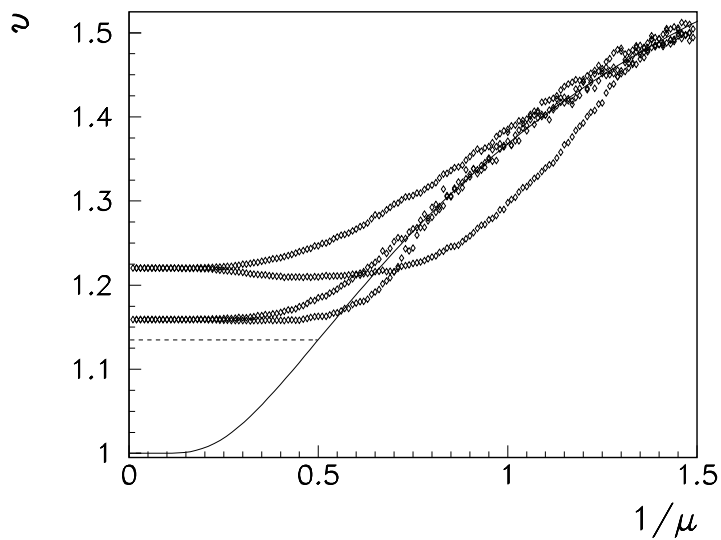}}
\caption{Hysteresis cycles in a ``cooling-heating'' experiment. 
The lower branch of the cycles represents the decompression.
The annealing rates are (from top to bottom) 
$r = 1/3 \cdot 10^{-4}, 10^{-6}$. 
}
\label{isteresi}
\end{center}
\end{figure}

Once we obtained the experimental  equation of state,
we can evaluate the entropy variation  
of the reservoir by numerical integration, which is given by:
\begin{eqnarray}
S(\mu_{\rm f})= S(\mu_{\rm i}) - \int_{\mu_{\rm i}}^{\mu_{\rm f}} \mu 
\frac{d\rho}{d\mu} d\mu \,\,.
\end{eqnarray}
This ``calorimetric entropy'' in presence of irreversible effects will 
be different from the thermodynamical entropy $S_{\rm  eq}$.
Indeed, in fig.~\ref{entropy} we see that when the relaxation time 
exceeds the inverse of the annealing rate the numerical data remain 
consistently above the equilibrium curve $S=S_{\rm eq}(\mu)$. 
If one were given only the {\em dynamical} data of 
figs.~\ref{annealing} and \ref{entropy}, one would feel tempted to
extrapolate the equilibrium specific volume and entropy to lower
 values of $1/\mu$.
Then, given that both these quantities are bounded, one could 
conclude that the Kauzmann ``temperature'', defined here as
\begin{eqnarray}
        \frac{1}{\mu_{\rm K}} 
        &=&
         \lim_{r \rightarrow 0} \frac{1}{\mu_{\rm g}(r)} \,, 
\end{eqnarray}
is different than zero and therefore that there {\em has} to be a 
static transition. 
This is the usual argument, known as Kauzmann's paradox, 
according to which the glassy state is related to 
the existence of a thermodynamic phase transition.
Of course, here there is no such static transition: 
in this simple case we have access to the whole equilibrium curves, 
which are perfectly analytical though they change concavity rather 
sharply.
Irreversibility effects are also evident when we let $1/\mu$ perform a 
cycle: in this case the specific volume appear to follow a hysteresis
loop whose area decreases as the compression speed 
decreases~(fig.~\ref{isteresi}).

\section{Structural relaxation}
        
We now turn to the behaviour of the system after a sudden
quench to a subcritical value $1/\mu< 1/\mu_{\rm c}$. 
In order to allow the system to reach more rapidly the 
asymptotic regime we perform a ``gentle'' quench i.e. 
starting from a configuration with density $0.75$ 
corresponding to a chemical potential closer to 
$\mu_{\rm c}$.
\begin{figure}
\begin{center}
\leavevmode
\epsfysize=230pt{\epsffile{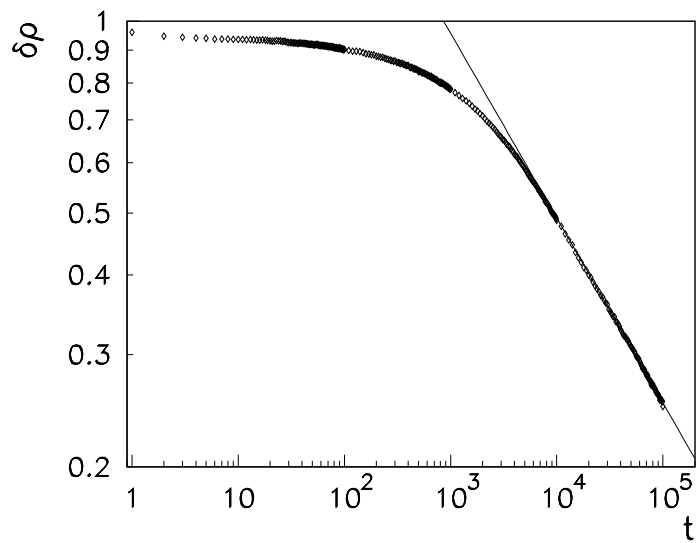}}
\caption{Time relaxation of the excess density 
after a subcritical quench,
$\delta \rho=(\rho_{\rm c}-\rho)/(\rho_{\rm c}-\rho_0)$.
The straight line is $\delta \rho \sim  t^{-z}$
with $z \simeq 0.3$.}
\label{relaxd}
\end{center}
\end{figure}
Figure~\ref{relaxd} shows the time relaxation of particles density
after a subcritical quench at $1/\mu= 1/2.2$. We see that $\rho$ never 
exceeds the threshold $\rho_{\rm c}$, but rather  approaches it like a 
power law in time:
\begin{eqnarray}
\rho_{\rm c}-\rho(t) &\sim& t^{-z} \,\,,
\end{eqnarray}
where $t$ is the time elapsed after the quench and where the  
exponent $z \simeq 0.3$.
Therefore the diffusion coefficient $D$ of particles after a 
subcritical quench vanishes as  
\begin{eqnarray}
D(t) &\sim& t^{-\zeta} \,\,,
\end{eqnarray}          
with the exponent $\zeta=z \phi$ quite close to one.
This is closely related to the aging behavior
observed in the two-time mean-squared displacement of particles,
$B(t,s)$. Indeed, in a simple minded approach such a quantity
would be given by
\begin{eqnarray}
B(t,s) &=& \int_{s}^{t} d\tau \, D( \tau) \,,
\end{eqnarray}     
from which follows the simple logarithmic aging
\begin{eqnarray}
B(t,s) & \sim &  \log (t/s) \,,
\end{eqnarray}    
in good agreement with the numerical results and the analytical
solution of the associated singular diffusion model~\cite{PeSe}.
Since the size of the system considered here is finite, equilibrium
will eventually be reached, (since almost any two allowed configurations
can be connected by a path of allowed moves), but with times which grow
fast as $ L \rightarrow \infty$.

\subsubsection*{Activated processes.}
\begin{figure}
\begin{center}
\leavevmode
\epsfysize=230pt{\epsffile{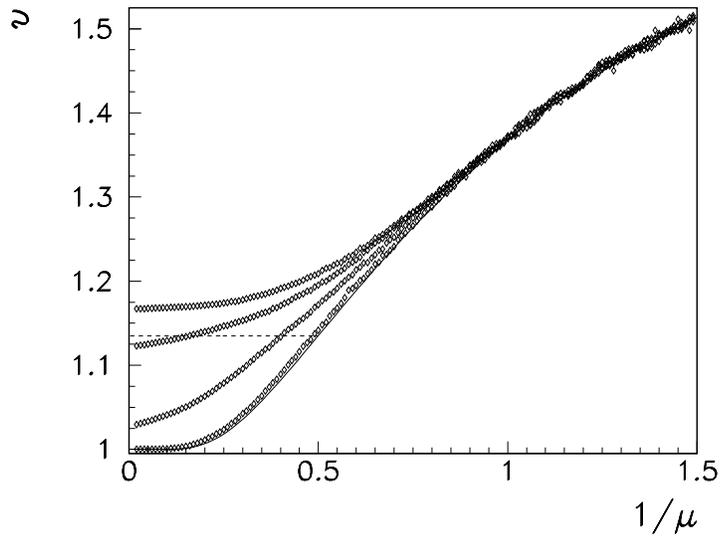}}
\caption{Compression experiment in presence of activated hopping processes
for a fixed annealing rate, $r = 10^{-5}$ and
different activation probabilities $p=10^{- k}$ with $k=1,\,2,\,3,\,4$.}
\label{ann_act}
\end{center}
\end{figure}

\begin{figure}
\begin{center}
\leavevmode
\epsfysize=230pt{\epsffile{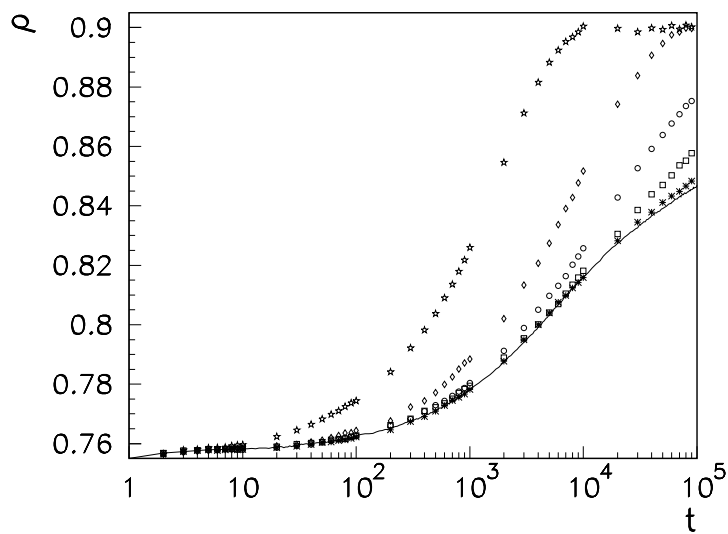}}
\caption{Density relaxation in presence
of activated hopping processes: 
 $p=10^{- k}$,  $k=1,\,2,\,3,\,4,\,5$ (from left to right); 
the continuous line represents the data with $p=0$.
The subcritical value of the quench, $1/\mu= 1/2.2$, 
corresponds to an equilibrium density $\rho \simeq 0.9$}
\label{d_act}
\end{center}
\end{figure}

It is interesting to investigate the role of activated hopping processes
in the low-temperature phase of glassy systems since they are 
responsible of restoring the ergodicity broken at the glass 
transition, and it is important to know the characteristic time scale
 on which this equilibration process takes place.
As pointed out in ref.~\cite{KoAn}, the activation processes can be 
simply implemented in this model by allowing the violation of the 
kinetic constraint with a given probability $p$.
Figure~\ref{ann_act} shows an example of a $v$ vs. $1/\mu$ plot in a 
compression experiment at fixed annealing rate for several values of 
the activation probabilities, $p$. 
As expected, in this case the system become able to cross the threshold and, 
after a certain value of $p$,  $p^*$, it follows the full equilibrium curve; 
we can also see that, for $p$ below $p^*$, the dependence of the relaxation 
time from $1/\mu$ is not affected by $p$, since the annealing curves depart 
from the equilibrium one approximately at the same point.  
The relation between the activation probability $p^*$ and
the equilibration time is better characterized by looking at behaviour of the
density after a  sudden quench.
In fig.~\ref{d_act} the relaxation curves in presence of activated 
processes are compared with that one obtained previously for $p=0$
(fig.~\ref{relaxd}). 
If we conventionally defined the ergodicity time, $\tau_{\rm erg}(p)$, 
as the time at which the curves with $p \neq 0$ depart from that  
one at $p=0$, it appears that this characteristic time follows a 
power-law, $\tau_{\rm erg} (p) \sim p^{-\alpha}$, with an exponent 
$\alpha \simeq 1$. 
A similar result was obtained by Castellano and Franz (unpublished).
This seems to provide a further evidence of the existence of a purely 
dynamical glass transition in this model.


\section{Conclusions}

To summarize we have shown that three dimensional lattice-gas 
models defined by short range kinetic constraint and trivial 
equilibrium Boltzmann-Gibbs measure display many features
of the fragile glass behaviour. 
Glassy phenomena may therefore have a purely dynamical or kinetic
origin unrelated to an underlying equilibrium phase transition, 
even in finite dimension and in absence of metastable states.
If this kinetic model exhibits a true dynamical transition
it would provide a microscopic realization, in finite dimension, 
of the mechanism invoked by the ideal mode-coupling theory for 
the glass transition.
Of course, it is hard to establish from numerical simulations
the existence of such a dynamical transition.
Indeed, a comparison with the backbone percolation problem show
that the linear size-dependence of the critical threshold cannot be 
faster than~\cite{KoAn}:
\begin{eqnarray}
1-\rho_{\rm c}(L) & \sim & 1/\log(\log L) \,,
\end{eqnarray}                                  
therefore even if $\lim_{L \rightarrow \infty } \rho_{\rm c}(L)=1$
(i.e. there is no ideal dynamical glass transition), 
the length-scale over which such a value would be observable is not 
experimentally accessible.
In this respect, the emergence in purely kinetic models of a well 
defined macroscopic effective temperature~\cite{CuKuPe} associated 
with the violation of the fluctuation-dissipation theorem appears 
quite surprising~\cite{Se}.
Indeed, given the non-holonomic nature of kinetic constraints and the trivial
Hamiltonian of the model, it would be interesting to understand whether a 
statistical mechanics approach based on the calculation of some restricted
partition function  is able to predict the features of the glassy
phase and in particular the value of the so called fluctuation-dissipation 
ratio.
A first step in this direction  would consist in to define a {\it kinetic}
analogue of the metastable states by considering, for example, as
metastable those system configurations where all particles are 
blocked by the kinetic constraint; and then find a way to count them.

\ack

I warmly thank J.~Kurchan and L.~Peliti for the collaboration
leading to the results presented here.
I also thank L.~Berthier, \'A.~de~Campos, S.~Franz and W.~Kob 
for interesting discussions.
This work is supported by the contract ERBFMBICT983561.

\end{document}